\documentstyle[12pt]{article}

\newcommand{\rbox}[1]{\vbox{\hrule height.8pt%
                \hbox{\vrule width.8pt\kern5pt
                \vbox{\kern5pt\hbox{#1}\kern5pt}\kern5pt
                \vrule width.8pt}
                \hrule height.8pt}}

\begin{document}

\begin{titlepage}

\begin{flushright}
KUNS-1371\\
HE(TH)95/21\\
hep-ph/?????????
\\
January, 1996
\end{flushright}

\vskip 0.35cm

\begin{center}
{\large \bf
Possible Candidates for SUSY E$_6$ GUT with an Intermediate Scale}

\vskip 1.2cm

Joe Sato\footnote
{e-mail: joe@gauge.scphys.kyoto-u.ac.jp}
\vskip 0.4cm

{\it Department of Physics, Kyoto University,\\
      Kyoto 606-01, Japan}

\begin{abstract}
We study the possibility of an intermediate scale existing in
supersymmetric E$_6$ grand unified theories. The intermediate scale
is demanded to be around $10^{12}$ GeV so that
neutrinos can obtain masses
suitable for explaining the
experimental data on the deficit
of solar neutrinos with
the Mikheev-Smirnov-Wolfenstein solution and the existence
of hot dark matter.
We require that at the intermediate scale, a certain
symmetry breakdown to the Standard Model symmetry.
We show that only a few E$_6$ subgroups are likely
to be realized as the intermediate symmetry, though there are
many candidates for the intermediate symmetry in E$_6$ GUT.
\end{abstract}

\end{center}
\end{titlepage}

\section{Introduction}
When we construct a Grand Unified Theory(GUT) based on
SO(10)\cite{SO10} and E$_6$\cite{E6},
in general, we have many extra fields which are contained in the same
multiplets as those of the quarks and leptons.
Under the Standard Model(SM) they can have mass terms
because they belong to a real representation under
the SM symmetry: some of them are singlet fermions under the SM
and the others appear
with their complex conjugate representation.
Singlet fermions may play the role of right-handed neutrinos.
Then the scale of the extra fields
is expected
to be a scale below which the SM is realized\footnote{
In the case that E$_6$ breaks down to the SM symmetry with several
scales, the scale of the extra fields may be higher than the scale
below which the SM is realized.}.

It is well known that in the Minimal Supersymmetric
Standard Model (MSSM) the present experimental values
of gauge couplings are successfully unified at a unification scale
$M_U \simeq 10^{16}$GeV \cite{Amal}. This fact implies
that if we would like to consider the gauge unification,
it is favorable that the symmetry of the GUT breaks down
to that of the SM at the unification scale.
In this case the scale of the right-handed neutrinos
$M_{\nu_R}$ and that of the other extra fields are
expected to be the unification scale $M_U$. This means also
that there is no intermediate scale between
the supersymmetry(SUSY) breaking scale and the unification scale.

On the other hand
it is said that $M_{\nu_R} \sim 10^{10-12} {\rm GeV}$\cite{yana}.
The experimental data on the deficit
of the solar neutrino can be explained by
the Mikheyev-Smirnov-Wolfenstein(MSW) solution \cite{WMS}.
According to one of the MSW solutions,
the mass of the muon neutrino seems to be
$m_{ \nu_\mu}\simeq10^{-3}$ eV.
Such a small mass can occur by the seesaw
mechanism \cite{seesaw}: a muon neutrino can
acquire a mass of O($10^{-3}$) eV
if the Majorana mass of the right-handed muon neutrino
is about $10^{12}$ GeV.

Then how can the right-handed neutrinos
acquire masses of  about $10^{12}$ GeV
when we have no scale other than $M_U$?
There are
several possibilities for the right-handed neutrinos to obtain
masses of the intermediate scale, $M_{\nu_R}\simeq10^{12}$ GeV.
First, radiative
correction of GUT scale physics, what we call
the Witten mechanism \cite{witten},
can induce $M_{\nu_R}$.
In a supersymmetric model, however, this
mechanism cannot work because the
non-renormalization theorem \cite{nonreno} protects
inducement of terms via radiative corrections which are not contained
in the original
Lagrangian. The second possibility is that the Yukawa coupling of right-handed
Majorana neutrino is so small that the mass may be the
intermediate scale even if it originates at the GUT scale. Third, singlet
Higgs particles develop a vacuum expectation value at the
intermediate scale to supply the mass of
$M_{\nu_R}$ to $\nu_R$.
In unrenormalizable models such as supergravity
these latter two possibilities may be realized.

Our point of view is, however, that it is more natural to consider
that one energy scale corresponds to a dynamical phenomenon,
for instance a symmetry breaking.
Thus we are led to another possibility: that a certain
intermediate group breaks
down to the standard group at the intermediate scale at which right-handed
neutrinos gain mass.
This idea is consistent with the survival hypothesis.
In previous papers we examined whether
it is possible to have an intermediate symmetry
in a SUSY SO(10) GUT\cite{BST,Joe}.
We saw that there is a possibility to have a SUSY SO(10) GUT
with an intermediate symmetry
${\rm SU(2)}_L \times {\rm SU(2)}_R \times {\rm SU(3)}_C \times
{\rm U(1)}_{B-L}$\cite{BST}
and actually we can construct such a SUSY SO(10) GUT\cite{Joe}.

In this paper we consider E$_6$ GUT as an extension
of SO(10) GUT. In E$_6$ GUT one family is embedded into one
irreducible multiplet.
We examine whether we can have
an intermediate symmetry in a SUSY E$_6$ GUT
using the same method of ref. \cite{BST}.

First we show what groups can be the intermediate symmetry.
Next we give a brief review of the method of ref. \cite{BST}.
Then we give the result.
Finally we give a summary and discussion.

\section{Intermediate Group}

\subsection{Matter content}
One quark and lepton family is embedded in E$_6$ {\bf 27}.
Then there are 12 extra matter fields in {\bf 27}.
According to their quantum number we denote them as follows.
There are two SM singlets. We label them $\nu_R$.
When we need to distinguish them from each other, we use a subscript
1 or 2. Others have the same quantum numbers as those of
the right-handed
down type quark $d^c$ and its charge conjugate $\overline{d^c}$
and those of lepton doublet $l$ and its charge conjugate
$\overline l$. The definition of notation for quarks and leptons
is as follows.

\begin{eqnarray*}
\begin{tabular}{lll}
$E_6$&
SU(2)$_L \times$ SU(3)$_C \times$ U(1)$_Y$ &\\
27&(2,3,1/6)&$q$\\
&(1,$\overline 3$,-2/3)&$u^c$\\
&2$\times$(1,$\overline 3$,1/3)&$d^c_1,d^c_2$\\
&2$\times$(2,1,-1/2)&$l_1,l_2$\\
& (1,1,1)&$e^c$\\
&(1,3,-1/3)&$\overline{d^c}$\\
& (2,1,1/2)&$ \overline l$\\
&2$\times$(1,1,0)&$\nu_{R1},\nu_{R2}$
\end{tabular}
\end{eqnarray*}

\subsection{Intermediate Group}
We have many kinds of $G_{\rm intermediate}$.
We list all of them and show the assignment of quantum numbers.
We refer to the regular maximal subgroup of E$_6$ to find
$G_{\rm intermediate}$. To know the
decomposition of representations and subgroups, see  ref.\cite{slan}.
There are three regular maximal subgroups in E$_6$:

\begin{eqnarray*}
{\rm A)}&&{\rm SO(10)} \times {\rm U(1)}\\
{\rm B)}&&{\rm SU(3)} \times {\rm SU(3)} \times {\rm SU(3)}\\
{\rm C)}&& {\rm SU(2)} \times {\rm SU(6)}.
\end{eqnarray*}

$G_{\rm intermediate}$ must consist of at least three
simple groups because at the intermediate scale
none of the three gauge couplings coincide with the other.
Moreover, one of the simple groups must contain SU(3)$_C$, the color group
and one of the other must contain SU(2)$_L$, the weak group.

\vspace{0.5cm}
A) Subgroup contained in SO(10) $\times$ U(1)$_X$

Under SO(10) $\times$ U(1)$_X$, E$_6$ {\bf 27}
becomes 1(4) + 10(-2) + 16(1).

By the reason mentioned above,
we have to consider the subgroups of SO(10):
SO(10) $\times$ U(1) $\supset
{\rm SU(2)} \times {\rm SU(2)} \times {\rm SU(4)} \times {\rm U(1)}$.
Under this subgroup,
E$_6$ {\bf 27} is decomposed to be (1,1,1)(4) + (2,2,1)(-2) +
(1,1,6)(-2) + (2,1,4)(1) + (1,2,$\overline 4$)(1).
One of the SU(2)'s must be identified with
SU(2)$_L$. We can give two ways of meaning
to the other SU(2) according to the definition
of the hypercharge: i)One is what we call
SU(2)$_R$.
 ii) The other is the diagonal group to the SM group.

Thus we see subgroups as follows:
\begin{eqnarray*}
{\rm i)} &
1)
&{\rm SU(2)}_L \times {\rm SU(2)}_R \times {\rm SU(4)}_{PS}
 \times {\rm U(1)}_X,\\
 &2)&{\rm SU(2)}_L \times {\rm U(1)}_R \times {\rm SU(4)}_{PS}
 \times {\rm U(1)}_X,\\
 &3)&{\rm SU(2)}_L \times {\rm SU(2)}_R \times {\rm SU(3)}_C
\times {\rm U(1)}_{B-L}
\times {\rm U(1)}_X\\
{\rm ii)}&
4)& {\rm SU(2)}_L \times {\rm SU(4)}_{PS'}
 \times {\rm U(1)}_X \times G_4 \\
&& G_4 = \hbox{\  SU(2) and its subgroups}.
\end{eqnarray*}

In the case i) the U(1)$_X$ is irrelevant to
the SM symmetry. Quarks and leptons are contained in SO(10) {\bf 16}.
The subgroup is recognized as a direct product of
Pati-Salam group (${\rm SU(2)}_L \times {\rm SU(2)}_R
\times {\rm SU(4)}_{PS}$) \cite{Pati} and U(1)$_X$.
There are also subgroups without U(1)$_X$.
Such  subgroups are SO(10) subgroups and hence
when one of those subgroups is realized as the intermediate
symmetry, we cannot see any difference from SO(10) GUT\cite{BST}.
We will not consider these three groups in this paper.
The groups 1) - 3) have U(1)$_X$, a reflection of E$_6$ GUT.

In the case ii) the
hypercharge Y is given by 1/4 X - 1/12 PS',
where PS' is SU(4) $T^{15}$ component charge. $T^{15}_4$ = diag\{1,1,1,-3\}.
The way of embedding quark and lepton is as follows:

\begin{eqnarray*}
\begin{tabular}{lll}
SU(2)$_L \times$ SU(4)$_{PS'} \times$ U(1)$_X$&
SU(2)$_L \times$ SU(3)$_C \times$ U(1)$_Y$ &\\
(1,1,4)& (1,1,1)&$e^c$\\
(2,1,-2)&(2,1,-1/2)&$l$\\
(1,6,-2)& (1,$\overline 3$,-2/3) + (1,3,-1/3)&$u^c ,\overline{d^c}$\\
(2,4,1)&(2,3,1/6) + (2,1,1/2)&$q , \overline l$\\
(1,$\overline 4$,1)&(1,$\overline 3$,-2/3) +(1,1,0)& $d^c$ , $\nu_R$
\end{tabular}
\end{eqnarray*}

There are other ii-type subgroups with a form of ``the SM symmetry''
$\times$ $G$ where $G$ are SU(2) $\times$ U(1) and its subgroups.
In the case that one of those groups is
realized as the intermediate group, we don't have any constraint
from the unification condition. The reason is that though
there are extra multiplets contained {\bf 27} of E$_6$
in the intermediate region, these
multiplets are identified with {\bf 10} of SO(10), an irreducible
representation of a simple group.
When we add a full multiplet of a simple group like SO(10),
the prediction of the gauge unification by MSSM is not spoiled.
Thus we don't consider these subgroups here.

\vspace{0.5cm}
B) Subgroups contained in SU(3) $\times$ SU(3) $\times$ SU(3)

Under  SU(3)$\times$ SU(3) $\times$ SU(3),
E$_6$ {\bf 27} becomes ($\overline 3$,3,1) + (3,1,3) +
(1,$\overline 3,\overline 3$).

In this case one of the SU(3)'s is identified with SU(3)$_C$
and one of the others is identified with SU(3)$_L$,
the group containing SU(2)$_L$.
We have following subgroups here:\footnote{
Exactly, we have the subgroup 3) as the subgroup of
SU(3) $\times$ SU(3) $\times$ SU(3). We omit such a duplication
in the following.}

\begin{eqnarray*}
&5)&{\rm SU(3)}_L \times {\rm SU(3)}_R \times {\rm SU(3)}_C,\\
 &6)&{\rm SU(2)}_L \times {\rm SU(3)}_R \times {\rm SU(3)}_C
 \times {\rm U(1)}_Z,\\
 &7)&{\rm SU(3)}_L \times {\rm SU(2)}_R \times {\rm SU(3)}_C
 \times {\rm U(1)}_{Z'},\\
 &8)&{\rm SU(3)}_L \times {\rm U(1)}_{R'} \times {\rm SU(3)}_C
 \times G_8,\qquad G_8 = \hbox{\ SU(2) and its subgroups}.\\
\end{eqnarray*}
The hypercharge Y is given by 1/6 Z -1/2 $T^3_R$ + 1/6 Z',
where Z and Z'
are SU(3) $T^8$ component charge. $T^8_3$ = diag\{1,1,-2\}.
$T^3_R$ is an SU(2)$_R$ generator and defined by diag\{1,-1\}.
R' is given by  3/2 $T^3_R$ - 1/2 Z.
The way of embedding quark and lepton is as follows:

\begin{eqnarray*}
\begin{tabular}{lll}
SU(3)$_L \times$ SU(3)$ \times$ SU(3)$_C$&
SU(2)$_L \times$ SU(3)$_C \times$ U(1)$_Y$ &\\
($\overline 3$,3,1)& 2 $\times$ (2,1,-1/2) +  (2,1,1/2) +
2 $\times$ (1,1,0) + (1,1,1)&
$l ,\overline l,\nu_R,e^c$\\
(3,1,3)&(2,3,1/6) + (1,3,-1/3) & $q,\overline{d^c}$\\
(1,$\overline 3,\overline 3$)&
 (1,$\overline 3$,-2/3) + 2 $\times$ (1,3,1/3)&$u^c, d^c$
\end{tabular}
\end{eqnarray*}

\vspace{0.5cm}
C) Subgroups contained in SU(2) $\times$ SU(6)

First we decompose SU(6) into its subgroup because,
as mentioned above, there must exist at least three
simple groups.
\begin{eqnarray*}
  {\rm SU(6)} \supset
\left\{
  \begin{array}{l}
{\rm SU(5)} \times {\rm U(1)},\\
{\rm SU(2)} \times {\rm SU(4)} \times {\rm U(1)},\\
{\rm SU(3)} \times {\rm SU(3)} \times{\rm U(1)}.
  \end{array}
\right.
\end{eqnarray*}
The second group and its subgroups are same as that appearing
in A) (SO(10)case). The third one coincides with B) case.

Then as the subgroup of SU(6) we consider only SU(5) $\times$
U(1). As mentioned above, the SU(2) is identified with SU(2)$_L$.
SU(5) is identified with SU(5)$_C$, a group which contains
SU(3) color group. Under SU(2)$_L \times$ SU(5)$_C \times$
U(1), {\bf 27} is decomposed to (2,$\overline 5$,-1) + (2,1,5)
+ (1,5,-4) + (1,10,2).
We can give two ways of meaning on SU(5)$_C$
according to the definition of the hypercharge.
\begin{eqnarray*}
9)&&{\rm SU(2)}_L \times {\rm SU(5)}_C \times {\rm U(1)}_W,\\
10)&&{\rm SU(2)}_L \times {\rm SU(5)}_{C'} \times {\rm U(1)}_W.
\end{eqnarray*}
To give hypercharge first we see
SU(2)$_R \times$ SU(3)$_C \times$ U(1)$_X$,
the subgroup of SU(5). Under it SU(5) {\bf 5} is (2,1,3) + (1,3,-2).
Then the hypercharge Y is equal to
2/15 X +1/10 W in 9) and 1/30 X + 1/2 $T^3_R$ - 1/10 W' in 10).
The way of embedding the quarks and leptons is as follows:\footnote{
In the following table the definition of {\bf 3} and {\bf $\overline 3$}
is opposite to the standard one.}

\begin{eqnarray*}
\begin{tabular}{llll}
9)&SU(2)$_L \times$ SU(5)$_C \times$ U(1)$_W$&
SU(2)$_L \times$ SU(3)$_C \times$ U(1)$_Y$ &\\
&(2,$\overline 5$,-1)& 2 $\times$ (2,1,-1/2) + (2,,$\overline 3$,1/6)
&$l ,q$\\
&(2,1,5)&
(2,1,1/2)&$\overline l$\\
&(1,5,-4)&2 $\times$ (1,1,0) + (1,3,-2/3)&$\nu_R,u^c$\\
&(1,10,2)&(1,1,1) + (1,$\overline 3$,-1/3) + 2 $\times$
(1,3,1/3)& $e^c, \overline{d^c},d^c$
\\
\end{tabular}
\end{eqnarray*}

\begin{eqnarray*}
\begin{tabular}{llll}
10)&SU(2)$_L \times$ SU(5)$_C \times$ U(1)$_W$&
SU(2)$_L \times$ SU(3)$_C \times$ U(1)$_Y$ &\\
&(2,$\overline 5$,-1)& (2,1,$\pm$1/2) + (2,,$\overline 3$,1/6)
&$\overline l,l ,q$\\
&(2,1,5)&
(2,1,1/2)&$l$\\
&(1,5,-4)&(1,1,1) + (1,1,0) + (1,3,1/3)&$e^c,\nu_R,d^c$\\
&(1,10,2)&(1,1,0) + (1,$\overline 3$,-1/3) + (1,3,-2/3) +
(1,3,1/3)& $\nu_R, \overline{d^c},u^c,d^c$
\\
\end{tabular}
\end{eqnarray*}

Their subgroups are the same as the groups appearing
in A) and omitted here.
Thus as the intermediate group we must consider these
10 subgroups.

\section{Unification Condition}

Here we give a brief review of ref.\cite{BST}.

The outline of our scenario is
\begin{eqnarray*}
{\rm E}_6 \longrightarrow G_{\rm intermediate}\longrightarrow {\rm MSSM},
\end{eqnarray*}
where $G_{\rm intermediate}$ is a subgroup of E$_6$ and contains
the SM symmetry. One quark and lepton family
are contained in E$_6$ {\bf 27}.

We require that there must exist multiplets
which can give mass to the right-handed neutrino
and also, if they exist,
to the other extra fields at the intermediate scale
where the intermediate symmetry breaks down to the SM.
In general, by introducing such multiplets
the gauge unification by MSSM is spoiled.
We cannot achieve the gauge unification without
introducing several multiplets at the intermediate region
between the GUT scale and the intermediate scale,
in addition to ordinary matter, three generations of quarks and
leptons and a pair of so-called Higgs doublets.

What condition is required of the multiplets at the
intermediate scale to recover the gauge unification?
In the following we make an analysis based on the RGE up to one loop.
We show the condition on the beta functions at the intermediate
scale in order to achieve the unification of gauge couplings.
The conditions of the unification are described by
\begin{eqnarray}
\alpha^{-1}_{Y}(M_S) &=& \alpha^{-1}_{U}(M_U) + \frac{1}{2\pi} b_Y R
                            + \frac{1}{2\pi} b'_Y (U - R),
\nonumber\\
\alpha^{-1}_{L}(M_S) &=& \alpha^{-1}_{U}(M_U) + \frac{1}{2\pi} b_{L} R
                            + \frac{1}{2\pi} b'_{L} (U - R),
\\
\alpha^{-1}_{C}(M_S) &=& \alpha^{-1}_{U}(M_U) + \frac{1}{2\pi} b_{C} R
                            + \frac{1}{2\pi} b'_{C} (U - R),
\nonumber
\end{eqnarray}
$b_i (i=Y,L,C)$'s with dash and without dash denote the
beta function in the lower scale and higher scale than the
intermediate scale $M_{\nu_R}$, respectively. $M_S$ is a certain scale
which is usually taken to be the SUSY breaking scale.
{\it R} and {\it U} are defined by
\begin{eqnarray}
R={\rm ln} \frac{M_{\nu_R}}{M_S},
\hspace{3em}
U={\rm ln} \frac{M_U}{M_S}.
\end{eqnarray}
These equations lead to the relation which {\it R} and {\it U} must satisfy,
\begin{eqnarray}
(b_Y - b_{L}) R + (b'_Y - b'_{L}) (U-R)
&=&
2\pi\left(\alpha^{-1}_{Y}(M_S) - \alpha^{-1}_{L}(M_S)\right),
\nonumber\\
(b_{C} - b_{L}) R + (b'_{C} - b'_{L}) (U-R)
&=&
2\pi\left(\alpha^{-1}_{C}(M_S) - \alpha^{-1}_{L}(M_S)\right).
\label{eq:ur}
\end{eqnarray}
Here we have assumed that in the lower scale MSSM is realized,
so the equation (\ref{eq:ur}) always has a solution  $U=R
$, which corresponds
to the case where there is  no intermediate scale physics.
Therefore if there is a nontrivial
intermediate scale $R$, the beta
functions must satisfy the following condition,
\begin{eqnarray}
(b_Y - b_{L})(b'_{C} - b'_{L}) -
(b_{C} - b_{L})(b'_Y - b'_{L}) = 0.
\end{eqnarray}
Since the beta functions in the MSSM are given by
\begin{eqnarray}
b_Y = \frac{33}{5},\ \ \ b_{L} = 1,\ \ \ b_{C} = -3,
\end{eqnarray}
the beta functions between the intermediate scale $M_{\nu_R}$ and GUT
scale $M_U$ must satisfy the equation,
\begin{eqnarray}
5 b'_Y - 12 b'_{L} + 7 b'_{C} = 0. \label{eq:UC}
\end{eqnarray}
which we call ``the unification condition''.\footnote{Though $b'_Y = b'_L
= b'_C$ satisfies the
unification condition, in this case the condition that
all couplings are unified is not fulfilled. Therefore this
case is excluded.}
This is a necessary condition on the gauge coupling unification
under the assumption that MSSM is realized in the lower scale.
When the equation (\ref{eq:UC}) is satisfied, R becomes an arbitrary
parameter. Therefore we introduce an intermediate scale $M_{\nu_R}$ as an
input parameter.

Using the unification condition for the
beta functions we make an analysis as follows:
Taking one combination of matter
content on the intermediate physics, we see whether the unification
condition is fulfilled or not\footnote{
Candidates for the matter contents in the intermediate
region are multiplets
included in representations {\bf 27, 78, 351, 351'} and {\bf 650}
of E$_6$. This is just an assumption. In general,
however, in the models mentioned above
it is difficult to include
a representation which is contained
in a higher representation of E$_6$ only. See the  statement
at the end of this section.}.
If this is the case,
we can calculate the unified scale $M_U$
and the gauge coupling $\alpha_U(M_U)$
at the unified scale using following equations,
\begin{eqnarray}
M_U \hspace{1em}&=& M_{\nu_R}\, {\rm exp}\left(2 \pi \frac{\alpha^{-1}_Y
                       (M_{\nu_R})-\alpha^{-1}_{L}(M_{\nu_R})}
                        {b'_Y - b'_{L}}\right),
\nonumber\\
\alpha_U(M_U) &=& \left(\alpha^{-1}_{L}(M_{\nu_R}) - \frac{1}{2\pi}
                b'_{L}(U-R) \right)^{-1},
\end{eqnarray}
once $M_R$ and $\alpha^{-1}_i(M_{\nu_R})$'s are given.

In principle we can calculate $\alpha^{-1}_i(M_{\nu_R})$'s from low-energy
experimental values of  $\alpha^{-1}_i$'s according to the RGE.
We choose, however, another
way to calculate  $\alpha^{-1}_i(M_{\nu_R})$ in order
to avoid ambiguities such as the SUSY breaking scale
$M_S$, the strong coupling $\alpha_{C}$, and so on.
Because we already know the unification scale $M_U^{\rm MSSM}$ and
${\alpha^{\rm MSSM}_U}^{-1}(M_U^{\rm MSSM})$
in MSSM GUT and above the intermediate
scale considered in this paper all couplings  $\alpha_i$'s
are small enough for one-loop approximation of RGE to work well,
we calculate $\alpha^{-1}_i(M_{\nu_R})$'s from the
input parameter ${\alpha^{\rm MSSM}_U}^{-1}(M_U^{\rm MSSM})$ at
the GUT scale $M_U^{\rm MSSM}$.
We choose input parameters from ref.\cite{Amal} as follows,
\begin{eqnarray}
M_U^{\rm MSSM} = 10^{16.3}\, {\rm GeV},
\ \ {\alpha^{\rm MSSM}_U}^{-1}(M_U^{\rm MSSM})=25.7. \label{eq:Input}
\end{eqnarray}

Then we select the matter
content which satisfies the following phenomenological criteria:
\begin{enumerate}
\item The unified scale $M_U$ is larger than $10^{16}$ GeV. This is
necessary for suppression of proton decay \cite{proton}.
\item The intermediate scale is taken at $10^{10}$, $10^{11}$,
$10^{12}$, $10^{13}$ or $10^{14}$ GeV because of right-handed
neutrino masses.
\item Any colored Higgs is not contained in the intermediate physics. This
is needed also for suppression of proton decay \cite{proton}.
\end{enumerate}

A search was made for all
possible combinations of matter contents by using a computer. This is
actually possible because the number of each matter multiplet
which we can take into account simultaneously cannot be
very large due to the
conditions we have already mentioned;
generally, the larger the number of matter contents is, the bigger
their contributions to beta functions are and the stronger the corresponding
couplings
$\alpha_i$'s become. As a result $\alpha_U^{-1}(M_U)$ becomes negative
below the unification scale $M_U$.

\section{Result}
Here we list the results in order.
As matter contents
which satisfies the unification condition eq.(\ref{eq:UC})
we show matter content which leads to the smallest unified coupling.

A) Subgroups of SO(10) $\times$ U(1)

In the case of i), the results are same
as those given in ref.\cite{BST} (SO(10) GUT case),
because extra U(1)$_X$ is orthogonal to the SO(10).
Of course, since there are extra multiplets in the intermediate region
as a result of E$_6$ GUT, the unified couplings are calculated to be
smaller.

When SU(4) is realized as the intermediate symmetry,
as case 1) and 2),
it is hard to have a solution.
To give mass to the right-handed neutrino we have to introduce
{\bf 10} of SU(4). Its contribution to the $\beta$ function
is so big that it
makes the inverse of the unification coupling,
$\alpha_U^{-1}(M_U)$, much smaller. The details are given in
ref.\cite{BST}.

In case 3) we have many solutions.
As an example we show a
result which gives the smallest value of
$\alpha_U \simeq 1/16.8$\footnote{
We can add an arbitrarily (1,1,1,0,$x$) + h.c, where $x$ is a charge of
U(1)$_X$. For example (1,1,1,0,4) is contained in E$_6$ {\bf 27}.
Such a multiplet is a singlet under the SM and hence
it does not contribute to the running of the gauge couplings.
Actually we need (1,1,1,0,4) + h.c in addition to (1,3,1,6,-2) + h.c
to break the intermediate symmetry down to the SM symmetry.
 }.
\begin{eqnarray}
&&\hspace{2em}
\alpha^{-1}_U(M_U) = 16.8,\ \ M_U=10^{16.3}\ {\rm GeV}
\ \ M_{\nu_R}=10^{12}\ {\rm GeV}
\hspace{5em}\nonumber\\
&& \underline{{\rm Higgs\ contents}}
\nonumber\\
&&\hspace{2em}
\begin{array}{lc}
(1,3,1,6,-2) + {\rm h.c} & 1 \\
(2,2,1,0,-2) + {\rm h.c}& 1 \\
(3,1,1,0,0) & 1 \\
(1,1,8,0,0) & 1
\end{array}
\nonumber
\end{eqnarray}
In the case of another
$M_{\nu_R}$,
$\alpha^{-1}_U(M_U)$ is slightly varied, though $M_U$ does not change.
The notation (1,3,1,6,-2) + h.c 1 indicates
that the representation
of the Higgs under the subgroup 3) is (1,3,1,6,-2) + h.c
and there is a single field.
(1,3,1,6,-2) + h.c is responsible for the right-handed neutrino mass.
(2,2,1,0,$\mp 2$) can be regarded as the standard Higgs in the MSSM.

In the  case of ii), as mentioned above, since
SU(4) is contained in the intermediate symmetry,
generally it is difficult to have a solution.
In other words the unified coupling is,
in general, calculated to be larger.
When $G_4$ is SU(2)\footnote{
We can add any number of (1,1,0,$n$) where $n$ is an arbitrary
representation of $G_4$ since it is a singlet under the SM
gauge group},
for example,
\begin{eqnarray}
&&\hspace{2em}
\alpha^{-1}_U(M_U) = 10.9,\ \ M_U=10^{16.3}\ {\rm GeV}
\ \ M_{\nu_R}=10^{12}\ {\rm GeV}
\hspace{5em}\nonumber\\
&& \underline{{\rm Higgs\ contents}}
\nonumber\\
&&\hspace{2em}
\begin{array}{lc}
(2,1,-2,2) + {\rm h.c} & 1 \\
(2,4,-1,1) + {\rm h.c}& 1 \\
(1,\overline 4,1,2) + {\rm h.c}& 1 \\
(1,\overline{10},2,1)  + {\rm h.c}& 1\\
(1,1,4,1)  + {\rm h.c}& 2\\
(3,1,0,1) & 1 \\
\end{array}
\nonumber
\end{eqnarray}
is a solution\footnote{
(2,1,-2,2) + {\rm h.c} and (1,1,4,1)  + {\rm h.c} + (3,1,0,1)
have same contribution to the unification condition and hence
we can replace them with each other.}.
$(1,\overline{10},2,1)  + {\rm h.c}$ gives mass
to right-handed neutrino.
$(1,\overline 4,1,2) + {\rm h.c}$ is responsible for mass terms
of $d^c \overline{d^c}$ and $l \overline l$.
(2,1,-2,2) + {\rm h.c} becomes the Higgs doublet for down type quarks
and leptons. (2,4,-1,1) + {\rm h.c} becomes the Higgs doublet for up type
quarks.

In the case that $G_4$ is U(1) or null
\begin{eqnarray}
&&\hspace{2em}
\alpha^{-1}_U(M_U) = 12.4,\ \ M_U=10^{16.3}\ {\rm GeV}
\ \ M_{\nu_R}=10^{12}\ {\rm GeV}
\hspace{5em}\nonumber\\
&& \underline{{\rm Higgs\ contents}}
\nonumber\\
&&\hspace{2em}
\begin{array}{lc}
(2,1,-2,1) + {\rm h.c} & 3 \\
(2,4,-1,0) + {\rm h.c}& 1 \\
(1,\overline 4,1,1) + {\rm h.c}& 1 \\
(1,\overline{10},2,0)  + {\rm h.c}& 1\\
(1,1,4,0)  + {\rm h.c}& 1\\
\end{array}
\nonumber
\end{eqnarray}
is a solution\footnote{Two (2,1,-2,1) + h.c
can be replaced by (1,1,4,0)  + h.c + (3,1,0,0)}.

B) Subgroups of SU(3) $\times$ SU(3) $\times$ SU(3)

There is no solution in chain 5).

In the case that 6) is realized as the intermediate symmetry
there are many solutions:
\begin{eqnarray}
&&\hspace{2em}
\alpha^{-1}_U(M_U) = 18..3,\ \ M_U=10^{16.3}\ {\rm GeV}
\ \ M_{\nu_R}=10^{12}\ {\rm GeV}
\hspace{5em}\nonumber\\
&& \underline{{\rm Higgs\ contents}}
\nonumber\\
&&\hspace{2em}
\begin{array}{lc}
(1,3,1,2) + {\rm h.c} & 2 \\
(2,3,1,-1) + {\rm h.c}& 1 \\
(1,3,1,-4) + {\rm h.c}& 1 \\
(1,1,3,4)  + {\rm h.c}& 2.
\end{array}
\nonumber
\end{eqnarray}
(2,3,1,-1) + {\rm h.c} become Higgs doublets
under the SM.
(1,3,1,2) + {\rm h.c} is needed for the mass term of
$d^c$ and $\overline{d^c}$. It is also necessary for $l \overline l$.
(1,3,1,-4) + {\rm h.c} gives mass to $\nu_R$'s.

There are 9 solutions which leads the unified gauge coupling
to 1/16.8 and 12 solutions corresponding to the unified
gauge coupling 1/15.3. Thus this breaking chain
seems to be hopeful.

Next we consider 7). There are many solutions though the unified coupling
becomes rather high:\footnote{
We can replace (1,3,1,0) with
(1,2,1,3) + h.c, (1,1,6,-4) + h.c with 2 $\times$ (1,3,1,0) +
(1,1,6,2) + h.c and so on.}

\begin{eqnarray}
&&\hspace{2em}
\alpha^{-1}_U(M_U) = 7.96,\ \ M_U=10^{16.3}\ {\rm GeV}
\ \ M_{\nu_R}=10^{12}\ {\rm GeV}
\hspace{5em}\nonumber\\
&& \underline{{\rm Higgs\ contents}}
\nonumber\\
&&\hspace{2em}
\begin{array}{lc}
(\overline 3,2,1,1) + {\rm h.c} & 1 \\
(\overline 3,1,1,-2) + {\rm h.c}& 1 \\
(6,2,1,1)+ {\rm h.c}& 1 \\
(1,3,1,0)& 1.\\
(1,1,8,0)& 1 \\
(1,1,\overline 3,-4) + {\rm h.c}& 1 \\
(1,1,6,-4)  + {\rm h.c}& 1.
\end{array}
\nonumber
\end{eqnarray}
($\overline 3$,2,1,1) + {\rm h.c} and ($\overline 3$,1,1,-2) + h.c
become Higgs doublets
under the SM.
They are also necessary for the mass term of
$d^c \overline{d^c}$ and $l \overline l$.
(6,2,1,1) + {\rm h.c} gives mass to the $\nu_R$'s.

Finally we give a result of 8). When $G_8$ is SU(2)\footnote{
We can add any number of (1,1,0,$n$) where $n$ is an arbitrary
representation of $G_8$ since it is singlet under the SM gauge group.},

\begin{eqnarray}
&&\hspace{2em}
\alpha^{-1}_U(M_U) = 12.4,\ \ M_U=10^{16.3}\ {\rm GeV}
\ \ M_{\nu_R}=10^{12}\ {\rm GeV}
\hspace{5em}\nonumber\\
&& \underline{{\rm Higgs\ contents}}
\nonumber\\
&&\hspace{2em}
\begin{array}{lc}
(\overline 3,1,1,2) + {\rm h.c} & 2 \\
(\overline 3,1,-2,1) + {\rm h.c}& 1 \\
(6,1,-2,1)+ {\rm h.c}& 1 \\
(1,8,0,1)& 2 \\
\end{array}
\nonumber
\end{eqnarray}
is a solution.
($\overline 3$,1,1,2) + {\rm h.c} and ($\overline 3$,1,-2,1) + h.c
become Higgs doublets
under the SM.
($\overline 3$,1,1,2) + {\rm h.c} plays the role of giving the mass
terms
$d^c \overline{d^c}$ and $l \overline l$.
(6,1,-2,1) + {\rm h.c} gives mass to the $\nu_R$'s.

If $G_8$ is U(1) or null, there are other solutions\footnote{
(3,$\overline 3$,1,1) + {\rm h.c} can be replaced with
(1,1,3,1) + h.c + (3,3,0,0) + h.c.}:

\begin{eqnarray}
&&\hspace{2em}
\alpha^{-1}_U(M_U) = 12.4,\ \ M_U=10^{16.3}\ {\rm GeV}
\ \ M_{\nu_R}=10^{12}\ {\rm GeV}
\hspace{5em}\nonumber\\
&& \underline{{\rm Higgs\ contents}}
\nonumber\\
&&\hspace{2em}
\begin{array}{lc}
(\overline 3,1,1,1) + {\rm h.c} & 1 \\
(\overline 3,1,-2,0) + {\rm h.c}& 1 \\
(6,1,-2,0)+ {\rm h.c}& 1 \\
(1,8,0,0)& 1 \\
(3,\overline 3,1,1) + {\rm h.c}& 1.
\end{array}
\nonumber
\end{eqnarray}

C) Subgroups of SU(2) $\times$ SU(6)

In the case 9) we have solutions, though
the unified coupling is too big to believe
that the subgroup 9) SU(2)$_L \times$ SU(5)$_C \times$ U(1)$_W$
is realized as the intermediate symmetry.

\begin{eqnarray}
&&\hspace{2em}
\alpha^{-1}_U(M_U) = 6.48,\ \ M_U=10^{16.3}\ {\rm GeV}
\ \ M_{\nu_R}=10^{12}\ {\rm GeV}
\hspace{5em}\nonumber\\
&& \underline{{\rm Higgs\ contents}}
\nonumber\\
&&\hspace{2em}
\begin{array}{lc}
(1,5,-4) + {\rm h.c} & 1 \\
(2,1,5) + {\rm h.c} &2\\
(2,\overline 5,-1) + {\rm h.c}& 1 \\
(1,\overline{10},8)+ {\rm h.c}& 1 \\
(3,1,0)& 2 \\
(1,24,0)& 2 \\
\end{array}
\nonumber
\end{eqnarray}
(1,5,-4) + {\rm h.c} plays the role of giving the mass term of
$d^c \overline{d^c}$ and $l \overline l$.
(2,$\overline 5$,-1) + {\rm h.c}
become the down type Higgs doublet
under the SM. (2,1,5) is the up type Higgs doublet.
(1,$\overline{10}$,8) + {\rm h.c} gives mass to the $\nu_R$'s.

On the other hand,
there is no solution in 10). The reason is that in this case
the representation of the necessary Higgs multiplets
becomes very high.

\section{Summary}

In most case we have solutions which satisfy the unification condition
eq.(\ref{eq:UC}). Among them, however, only four intermediate
groups can lead to a small unified coupling. 3)
${\rm SU(2)}_L \times {\rm SU(2)}_R \times {\rm SU(3)}_C
\times {\rm U(1)}_{B-L}
\times {\rm U(1)}_X$ and 6)
${\rm SU(2)}_L \times {\rm SU(3)}_R \times {\rm SU(3)}_C
\times {\rm U(1)}_Z$
seem possible to be realized as the intermediate group.
4) ${\rm SU(2)}_L \times {\rm SU(4)}_{PS'} \times {\rm U(1)}_X
\times G_4$  and 8)
${\rm SU(3)}_L \times {\rm SU(3)}_C
\times {\rm U(1)}_Z \times G_8$
may also be possible.

The reason why the unified coupling, in general,
becomes rather big is that there are extra multiplets
in the intermediate region which are contained in E$_6$ {\bf 27}.
By the intermediate symmetry they cannot acquire mass.
They contribute to the running of the gauge couplings.
The extra multiplets lead the unified gauge coupling
to larger value.

Thus in the case of E$_6$ GUT
it is difficult to have a solution.
We can
pick up favorable subgroups for the intermediate symmetry,
though there are many E$_6$ subgroups.

Since we have candidates for the intermediate symmetry
and the matter content in the intermediate scale,
it is possible to construct an E$_6$ GUT with an intermediate
scale. Even in E$_6$ GUT we can consider the right-handed
neutrino mass to be a reflection of symmetry breaking.

\begin{center}
 ACKNOWLEDGEMENTS
\end{center}

The author wish to acknowledge M.~Bando and
T.~Takahashi for valuable comments and discussion
and thanks A.~Bordner for proof reading.

\end{document}